\documentclass[preprint]{aastex}
\usepackage{mkfig}

\begin{document}
 
\title{\bf Molecular Hydrogen Emission Lines in {\em Far Ultraviolet
  Spectroscopic Explorer} Observations of Mira~B\altaffilmark{1}}

\author{Brian E. Wood\altaffilmark{2} and Margarita Karovska\altaffilmark{3}}

\altaffiltext{1}{Based on observations made with the NASA-CNES-CSA Far
  Ultraviolet Spectroscopic Explorer.  FUSE is operated for NASA by the
  Johns Hopkins University under NASA contract NAS5-32985.}
\altaffiltext{2}{JILA, University of Colorado and NIST, Boulder, CO
  80309-0440; woodb@origins.colorado.edu.}
\altaffiltext{3}{Smithsonian Astrophysical Observatory, 60 Garden St., 
  Cambridge, MA 02138; mkarovska@cfa.harvard.edu.}

\begin{abstract}

     We present new {\em Far Ultraviolet Spectroscopic Explorer} (FUSE)
observations of Mira~A's wind-accreting companion star, Mira~B.
We find that the strongest lines in the FUSE spectrum are H$_{2}$
lines fluoresced by H~I Ly$\alpha$.  A previously analyzed {\em Hubble
Space Telescope} (HST) spectrum also shows numerous Ly$\alpha$-fluoresced
H$_{2}$ lines.  The HST lines are all Lyman band lines, while
the FUSE H$_{2}$ lines are mostly Werner band lines, many of them never
before identified in an astrophysical spectrum.  We combine the FUSE and HST
data to refine estimates of the physical properties of the emitting H$_{2}$
gas.  We find that the emission can be reproduced by an H$_{2}$ layer with
a temperature and column density of $T=3900$~K and $\log N(H_{2})=17.1$,
respectively.  Another similarity between the HST and FUSE data, besides the
prevalence of H$_{2}$ emission, is the surprising weakness of the continuum
and high temperature emission lines, suggesting that accretion onto Mira~B
has weakened dramatically.  The UV fluxes observed by HST on 1999
August 2 were previously reported to be over an order of magnitude lower
than those observed by HST and the {\em International Ultraviolet
Explorer} (IUE) from 1979--1995.  Analysis of the FUSE data reveals
that Mira~B was still in a similarly low state on 2001 November 22.

\end{abstract}


\section{INTRODUCTION}

     Mira~A (o~Ceti, HD~14386) is the prototype for a class of pulsating
giant stars on the asymptotic giant branch.  The pulsations of Mira
variables help drive very strong winds from the surfaces of these stars.
Mass loss rate estimates for Mira~A itself generally fall in the range
$4\times 10^{-8}$ to $4\times 10^{-7}$ M$_{\odot}$ yr$^{-1}$
\citep{yy78,grk85,pfb88,pp90,grk98,nr01}.  Mira~A has a companion star,
Mira~B, which is located $0.6^{\prime\prime}$ away \citep{mk97},
corresponding to a projected distance of about 70~AU at its
distance of $128\pm 18$~pc \citep{macp97}.  Mira~A's wind is being
accreted by Mira~B, forming an accretion disk.  The Mira system is
attractive for studying accretion processes since it is one of the few
wind accretion systems in which the components of the system are
resolvable.

     Mira~B's accretion disk emits
broad, high temperature emission lines of C~IV $\lambda$1550,
Si~III] $\lambda$1892, and Mg~II $\lambda$2800, among others, which were
first observed by the {\em International Ultraviolet Explorer} (IUE)
\citep{dr85}.  The optical and UV continuum of Mira~B
appears to be dominated by the accretion based on its strong
variability on many timescales, and based on accretion rate estimates of
$(8-30)\times 10^{-10}$ M$_{\odot}$ yr$^{-1}$
\citep{bw72,yy77,mj84,dr85}.  Because of the complexities
involved with the wind accretion onto Mira~B, it is uncertain whether the
star is a white dwarf or a red dwarf.

     The UV spectrum of Mira~B was observed many times by IUE between 1979
and 1995 \citep{dr85}, and also by the Faint Object Camera
(FOC) instrument on the {\em Hubble Space Telescope} (HST) on 1995
December 11 \citep{mk97}.  The UV continuum and emission lines of
Mira~B show some modest variability within the 1979--1995 data, with fluxes
varying by about a factor of 2.  However, on 1999 August 2 the
Space Telescope Imaging Spectrograph (STIS) instrument on HST obtained a UV
spectrum from Mira~B that was radically different from any previous
observation \citep[][hereafter Paper I]{bew01}.  The fluxes of the
continuum and high temperature emission lines (e.g., C~IV $\lambda$1550,
Si~III] $\lambda$1892, Mg~II $\lambda$2800, etc.) were over an order
of magnitude lower than ever observed before.  Furthermore, the character
of the spectrum below 1700~\AA\ had changed dramatically, with the
spectrum dominated by many narrow H$_{2}$ lines rather than being
dominated by the aforementioned broad, high temperature lines.  These
H$_{2}$ lines are pumped by the strong H~I Ly$\alpha$ line, a fluorescence
mechanism that has been found to produce detectable H$_{2}$ emission
from the Sun and more recently from many other astrophysical sources
\citep{cj77,ab81,adm99,jav00,dra02}.
The surprising STIS data raise many new questions about the accretion onto
Mira~B.  Why did the UV fluxes fall so dramatically?  Where are all these
H$_{2}$ lines coming from?  Why were the H$_{2}$ lines not observed by IUE?

     \citet[][hereafter Paper II]{bew02} analyzed the H$_{2}$ lines in
detail.  They found that the dominance of H$_{2}$ emission in the 1999
HST/STIS data is due at least in part to an H~I Ly$\alpha$ line that is
{\em not} weaker than during the IUE era, unlike the continuum and
every other non-H$_{2}$ line in the spectrum.  Therefore, the H$_{2}$
lines pumped by Ly$\alpha$ appear much stronger relative to other UV
emission lines than before, when the H$_{2}$ lines were not even
detectable by IUE.  It was proposed in Papers I and II that the
fundamental cause of the change in Mira~B's UV spectrum was an order of
magnitude decrease in the accretion rate onto the star.  This
interpretation is supported by analysis of wind absorption in the
Mg~II h \& k lines at 2803 and 2796~\AA, respectively, which shows that
the accretion-driven mass loss rate from Mira~B at the time of the
HST/STIS observations is lower by about an order of magnitude from
what it was during the IUE era, consistent with the observed
decrease in accretion luminosity.  Exactly why the Ly$\alpha$ flux did
{\em not} decrease with everything else in the spectrum remains somewhat
of a mystery.  In Paper~II, we suggested that the Ly$\alpha$ emission may
have indeed decreased, but the weaker wind opacity at the time of
the HST/STIS observations allowed more Ly$\alpha$ emission to escape and
compensated for this decrease.

     As far as where the H$_{2}$ lines are coming from, several arguments
were presented in Paper II against the H$_{2}$ emission being from the
accretion disk.  Instead, the H$_{2}$ lines are most likely coming from
H$_{2}$ within Mira~A's wind, which is being heated and dissociated by
H~I Ly$\alpha$ as it approaches Mira~B.  The H$_{2}$ emission line ratios
and the amount of H$_{2}$ absorption observed for the pumping transitions
within Ly$\alpha$ are both consistent with an H$_{2}$ layer with
$T\approx 3600$~K and $\log N(H_{2})\approx 17.3$.  This temperature is
close to the dissociation temperature of H$_{2}$, suggesting that the
H$_{2}$ could be from an H$_{2}$ photodissociation front surrounding Mira~B.
A photodissociation front model presented in Paper II demonstrates that such
a front can indeed reproduce the properties of the H$_{2}$ emission,
although it was suggested that the collision of the winds of Mira~A and B
could also play a role in heating the H$_{2}$.  The H$_{2}$
photodissociation rate estimated from the data is roughly consistent with
Mira~B's $\sim 10^{-10}$ M$_{\odot}$~yr$^{-1}$ total accretion rate,
meaning that the H$_{2}$ we are seeing being fluoresced and dissociated by
Ly$\alpha$ is probably on its way to being ultimately accreted onto Mira~B.
Molecular hydrogen is the dominant constituent of Mira~A's wind by mass, so
the accretion processes relating to H$_{2}$ are particularly important.
The fluorescence, dissociation, and heating of the H$_{2}$ by Ly$\alpha$,
which is what the UV H$_{2}$ lines are probing, is therefore an important
step in the process of accretion onto Mira~B.

     In this paper, we report on new UV observations of Mira~B from the
{\em Far Ultraviolet Spectroscopic Explorer} (FUSE).  The FUSE satellite
observes the $905-1187$~\AA\ wavelength range, which is almost entirely
inaccessible to the HST.  The Mira binary system has never been observed in
this wavelength region, so the FUSE data allow us to search for new
emission line diagnostics for Mira~B.

\section{FUSE OBSERVATIONS}

     The FUSE satellite observed Mira~B on 2001 November 27 starting at
UT 9:04:57.  The observation consisted of 11 separate exposures through
the low resolution (LWRS) aperture, with a combined exposure time of
29,183 s.  The spectra were processed using version 2.0.5 of the
CALFUSE pipeline software.  Note that while Mira~A and Mira~B are both
within the $30^{\prime\prime}\times 30^{\prime\prime}$ LWRS aperture,
experience with IUE data demonstrates that Miras do not
produce any detectable emission below 2000~\AA\ \citep{bew00},
so any emission detected by FUSE should be from Mira~B.

     In order to fully cover its 905--1187~\AA\ spectral range, FUSE has a
multi-channel design --- two channels (LiF1 and LiF2) use Al+LiF coatings,
two channels (SiC1 and SiC2) use SiC coatings, and there are two different
detectors (A and B).  For a full description of the instrument, see
\citet{hwm00}.  With this design FUSE acquires spectra in 8 segments covering
different, overlapping wavelength ranges.  We coadded the individual
exposures to produce spectra for each segment.  We experimented with
cross-correlating the exposures before coaddition, but there are not many
spectral features strong enough to cross-correlate against, so the simple
straight coaddition proved more reliable.  Coaddition of the various
segments with each other can potentially lead to degradation of spectral
resolution, but we decided to do this anyway to increase the
low signal-to-noise (S/N) of our data.  We inspected the
individual segments carefully to ensure they were reasonably well
aligned before coadding them.

     Figure~1 shows two regions of the FUSE spectrum.  The top
panel is a coaddition of the SiC1B and SiC2A segments and the bottom
panel is a combination of the LiF1B and LiF2A segments.  We have rebinned
the two spectra by factors of 15 and 5, respectively, to increase
S/N.  The FUSE data highly oversample the line spread function, so this
rebinning does not severely degrade the spectral resolution even for the
factor-of-15 rebinning.  The bottom spectrum in Figure~1 has also been
smoothed slightly, but only for purposes of display.  No stellar emission
is detected between 990 and 1115~\AA, so that region of the FUSE spectrum
is not shown.

     The top panel of Figure~1 shows mostly only airglow lines of H~I
and O~I.  There is a broad emission feature at 976~\AA\ that may be
C~III 977.02~\AA\ emission from the star, but the feature is shifted from
its expected location by a suspiciously large amount
($\sim$ 200 km~s$^{-1}$),
making this identification tentative (see \S 5).  The bottom
panel shows numerous narrow lines that we identify
as H$_{2}$ emission, as will be described in detail below.  Many of these
lines are blended with the C~III $\lambda$1175 multiplet, making it unclear
whether C~III is really contributing any flux to the blend at all.
Both C~III lines shown in Figure~1 are strong lines frequently seen in
stellar spectra, and we expected to be able to detect these lines from
the Mira~B accretion disk.  We will discuss them further in \S5.

     We have fitted Gaussians to all of the stellar emission features shown
in Figure~1, using a chi-squared minimization routine to determine the
best fit \citep[e.g.,][]{prb92}.  The first three columns of Table~1 list
the fit parameters and their 1$\sigma$ uncertainties.  The parameters are
the central wavelength ($\lambda_{meas}$), flux ($f_{obs}$, in units of
$10^{-15}$ ergs~cm$^{-2}$~s$^{-1}$), and
full-width-at-half-maximum (FWHM).  The uncertainties are estimated using a
Monte Carlo technique whereby the fluxes in the spectrum are varied within
the bounds suggested by the flux error vector, and a large number of fits
are performed on the altered spectra to see how much fit parameters vary.
Our quoted 1$\sigma$ uncertainties are the 1$\sigma$ variations of the
Monte Carlo trials.  Figure~2 shows the fit to the complicated blend at
1176~\AA.  The narrow lines are all H$_{2}$ lines, and we assume the sum of
the two broad components is representative of the C~III emission.  The
parameters for both broad Gaussians are listed in Table~1 and tentatively
identified as C~III.

     The line spread function of FUSE is not particularly well
defined, so we have made no attempt to correct for instrumental broadening
in our fits.  The H$_{2}$ lines observed by HST/STIS have an average width
of $FWHM=19.7\pm 0.4$ km~s$^{-1}$, after correction for instrumental
broadening (Paper~II).  The effective resolution
of FUSE is $\sim 20$ km~s$^{-1}$, so we expect the FUSE H$_{2}$ lines
to have widths of $\sim 28$ km~s$^{-1}$.  This agrees reasonably well
with the measured line widths in Table~1, with blending perhaps being
responsible for the broadening of a few H$_{2}$ lines.  The H$_{2}$
lines have average redshifts of $+62$ km~s$^{-1}$ relative to their
rest wavelengths.  Considering the $\sim 5$ km~s$^{-1}$ uncertainty in the
FUSE wavelength calibration, this is consistent with the $+56.9\pm 0.2$
km~s$^{-1}$ velocity found for the STIS H$_{2}$ lines, which in turn is
consistent with the Mira system's radial velocity
of $\sim 56$ km~s$^{-1}$ \citep{pfb88,pp90,ej00}.

\section{IDENTIFYING THE H$_{2}$ LINES}

     Most of the H$_{2}$ lines listed in Table~1 have never before been
identified in any astrophysical spectrum.
In Paper~II, we found that all of the H$_{2}$ lines observed by HST/STIS
could be associated with Lyman band fluorescence sequences pumped by the
strong H~I Ly$\alpha$ line.  The Ly$\alpha$ emission excites H$_{2}$ from
various rovibrational states within the ground electronic state
($X^{1}\Sigma_{g}^{+}$) to the excited $B^{1}\Sigma_{u}^{+}$ state.
Radiative deexcitation back to various levels of the ground electronic
state yields the Lyman band fluorescence sequences observed by STIS.  We
conclusively identified H$_{2}$ lines in the STIS spectrum by finding many
lines in each fluorescence sequence, each having a Ly$\alpha$ pumping path,
and noting that the line ratios are at least roughly consistent
with the transition branching ratios.  This is harder to do in our
FUSE spectrum, because we have fewer lines to work with.  The
Lyman band sequences from Paper~II only contribute to a couple of the
H$_{2}$ lines in Figure~1 (at 1143~\AA\ and 1162~\AA).  The other narrow
lines are from previously unidentified sequences.

     We used the Lyman band line list of \citet{ha93a} to search
for possible matches to our narrow FUSE lines, but we quickly found
that most of these lines are clearly not Lyman band H$_{2}$.  However, we
were much more successful when we searched the Werner band H$_{2}$ line
list of \citet{ha93b}.  Just as emission from Ly$\alpha$ can
excite H$_{2}$ to the $B^{1}\Sigma_{u}^{+}$ state to produce Lyman band
emission, Ly$\alpha$ can also excite H$_{2}$ to the excited
$C^{1}\Pi_{u}$ state to produce Werner band emission, most of which
falls within the FUSE bandpass rather than the higher wavelength range of
STIS.  

     The Werner band fluorescence is of a slightly different character than
the Lyman band fluorescence.  Quantum selection rules require Lyman band
transitions to have $\Delta J\equiv J''-J'=\pm 1$, where $J'$ and $J''$ are
the upper and lower rotational quantum numbers, respectively.  Transitions
with $\Delta J=+1$ are defined as P-branch lines and those with
$\Delta J=-1$ are R-branch lines.  The $C^{1}\Pi_{u}$ state is actually
degenerate and consists of a $C^{+}$ state and a $C^{-}$ state.  The energy
levels of the two states are practically identical, but Werner band
transitions from $C^{+}$ must have $\Delta J=\pm 1$ (i.e., P and R branch
transitions), while transitions from $C^{-}$ must have $\Delta J=0$
(Q branch transitions).  The Werner band is therefore different from the
Lyman band in having a Q branch, but because the $C^{+}$ and $C^{-}$ states
are different the Q branch lines can only be pumped by Q branch transitions
within Ly$\alpha$, and the P and R branch lines can only be pumped by
P and R branch transitions.

     Perusal of the \citet{ha93a,ha93b} line lists leads to possible
H$_{2}$ identifications for all the narrow lines seen in the FUSE data,
but conclusive identification is complicated due to most potential new
fluorescence sequences having only 1 or 2 observed lines, and also due
to many of the lines possibly being blends of more than one H$_{2}$ line.
In order to clarify the situation and provide support for the detection of
some of the weaker H$_{2}$ lines, we perform a forward modeling
exercise using the results of Paper~II.  In Paper~II, we used the STIS
H$_{2}$ lines to estimate the temperature and column density of the
H$_{2}$, and we also reconstructed the Ly$\alpha$ profile that the H$_{2}$
must see in order to account for the observed H$_{2}$ flux.  Using these
results, we create a synthetic H$_{2}$ spectrum, considering all new
Lyman and Werner band fluorescence sequences that may be contributing to
the FUSE H$_{2}$ lines.  In Table~1, we list the line fluxes predicted
by this forward modeling exercise ($f_{mod}$).  In particular, Table~1
lists all lines that we believe contribute at least 10\% of the flux of an
observed FUSE H$_{2}$ feature based on the $f_{mod}$ values.  Note that
there is a broad, weak emission feature at 1134--1138~\AA\ that our forward
modeling exercise suggests may be partially H$_{2}$ emission.  However,
H$_{2}$ lines cannot seem to account for most of the flux of the line, so
we do not consider it a clearly detected H$_{2}$ feature.

     Another purpose of the forward modeling exercise, besides line
identification, is to see if the reconstructed Ly$\alpha$ profile and
H$_{2}$ parameters from Paper~II can reproduce the fluxes in our FUSE
data.  Table~1 shows that the predicted fluxes generally agree with the
measured fluxes rather well, with a few exceptions.  The Ly$\alpha$
profile must therefore be about the same at the time of the FUSE
observation (2001 November 27) as it was at the time of the STIS
observation (1999 August 2).  This means that we can consider the FUSE
and STIS H$_{2}$ lines together to refine the H$_{2}$ analysis presented
in Paper~II, which we will do in \S4.

     While trying to identify the FUSE H$_{2}$ lines, we also revisited
the STIS data and found one new STIS H$_{2}$ fluorescence sequence,
which is also listed in Table~1.  The five detected lines in this
sequence are individually rather weak, but collectively they amount to a
convincing detection.  This new sequence is important because it is
fluoresced at a lower wavelength than any other Mira~B H$_{2}$ sequence,
thereby providing a useful diagnostic for the H$_{2}$-observed Ly$\alpha$
flux at shorter wavelengths (see \S4).  We note that this sequence was
also detected in observations of the T~Tauri star TW~Hya \citep{gjh02,gjh03}.

     We emphasize that only a few of the H$_{2}$ lines listed in
Table~1 have been
identified in previous astrophysical spectra.  While fluoresced Lyman band
H$_{2}$ lines have been observed and analyzed from a number of different
sources (see \S 1), the list of Werner band detections is very short.
\citet{jdfb79} noted a couple Werner lines in solar spectra, which
are actually fluoresced by the O~VI $\lambda$1032 line rather than
Ly$\alpha$, and \citet{ew02} report FUSE detections of a few
Werner band H$_{2}$ lines from the prototype young accreting star system,
T~Tauri.  Planetary aurora produce abundant H$_{2}$ emission in both the
Lyman and Werner bands, but this is collisionally excited rather than
fluoresced emission \citep[e.g.,][]{pfm97}.

\section{COMPLETE ANALYSIS OF THE H$_{2}$ LINES OF MIRA B}

     Table~2 lists all the pumping transitions within Ly$\alpha$ that
are responsible for the H$_{2}$ lines observed by FUSE and STIS.  Our
analysis has expanded the list from the 13 Lyman band sequences analyzed in
Paper~II to 16 Lyman and 13 Werner sequences.  This increase in H$_{2}$
data is sufficient to justify a reanalysis of the H$_{2}$ lines to
see if conclusions from Paper~II remain unchanged.

     The ratios of H$_{2}$ fluxes within each fluorescence sequence are
inconsistent with the line branching ratios and are therefore clearly
affected by opacity effects (see Paper~II).  The opacity of the
H$_{2}$ lines depends on the level populations of the H$_{2}$ molecules
being fluoresced, which in turn depends on the temperature, T, and total
column density of the H$_{2}$, $N(H_{2})$.  The line ratios are therefore
a diagnostic for T and $N(H_{2})$.  In Paper~II, we developed a plane
parallel, Monte Carlo radiative transfer code to determine which T and
$N(H_{2})$ values best fit the line ratios.  We repeat this analysis
including the new H$_{2}$ data presented here.  For the blended lines in
Table~1, we divide up the line flux according to the percent contributions
to the line suggested by the $f_{mod}$ values.

     For each transition, we need to know the absorption strength ($f$),
and the energy ($E_{low}$) and statistical weight ($g_{low}$) of the lower
level.  The necessary atomic data are taken from \citet{ha93a,ha93b} and
\citet{id84}.  Table~2 lists the $f$,
$E_{low}$, and $g_{low}$ values of the Ly$\alpha$ pumping transitions.
Another factor considered in the radiative transfer model is dissociation.
Fluorescence to the $B^{1}\Sigma_{u}^{+}$ and $C^{1}\Pi_{u}$ states
can result in dissociation of the H$_{2}$ molecule rather than radiative
deexcitation back to the ground electronic state.  The dissociation
probabilities listed in \citet{ha00} are considered
in the model.  The $f_{dis}^{\prime}$ values in Table~2 indicate the
fractional probability of dissociation for each excitation to the
excited electronic state.  The $f_{dis}$ values indicate the fraction
of fluorescences within each sequence that ultimately lead to dissociation
rather than the emergence of an H$_{2}$ line photon, based on our
best-fit radiative transfer model (see below).  Line opacity
can cause multiple photoexcitations before an H$_{2}$ photon
emerges or an H$_{2}$ molecule is destroyed, so $f_{dis}>f_{dis}^{\prime}$.
In Paper~II, the \citet*{ha00} tables were read incorrectly,
resulting in the dissociation fractions being significantly overestimated.
The $f_{dis}^{\prime}$ and $f_{dis}$ values in Table~2 of Paper~II are
therefore too large in general.  Besides consideration of the new FUSE data,
a secondary reason for reanalyzing the H$_{2}$ lines is to revise our
analysis to see if conclusions made in Paper~II regarding the importance
of dissociation still stand.

     The only free parameters of the model and $T$ and $N(H_{2})$, and
Figure~3 shows $\chi^{2}_{\nu}$ contours measuring how well the
H$_{2}$ line ratios are reproduced by models with different T and
$\log N(H_{2})$ values.  The H$_{2}$ line ratios are best explained by an
H$_{2}$ layer with a temperature and column density of $T=3900$~K and
$\log N(H_{2})=17.1$, respectively, similar to the $T=3600$~K and
$\log N(H_{2})=17.3$ results from Paper~II.  Thus, neither the new FUSE
data nor the corrected treatment of dissociation change these values much.

     The third column of Table~2 lists the total Ly$\alpha$ flux
absorbed that emerges as H$_{2}$ line flux, $F_{obs}$, not including the
absorbed flux that leads to H$_{2}$ dissociation.  These $F_{obs}$
values include corrections for H$_{2}$ lines in each sequence that
are too weak to be detected, with the correction factors provided by the
best-fit radiative transfer model.  The $F_{obs}$ values in Table~2 are
slightly higher than those from Paper~II (for the 13 original Lyman band
sequences) due to the lower dissociation fractions.

     The $F_{obs}$ and $f_{dis}$ values in Table~2 can be used to
compute the Ly$\alpha$ flux that must be overlying each H$_{2}$
pumping transition within Ly$\alpha$, thereby allowing us to reconstruct
the Ly$\alpha$ profile seen by the H$_{2}$, $f_{0}(\lambda)$.  Once
again we refer the reader to Paper~II for details of this computation,
although here we assume our revised values of $T=3900$~K and
$\log N(H_{2})=17.1$ in the calculations.  Figure~4 shows the new
reconstructed Ly$\alpha$ profile.  The analysis assumes that the
fluoresced H$_{2}$ completely surrounds the star and that the
H$_{2}$ emission emerges isotropically.  If these assumptions are
inaccurate, the derived profile
will be different from the actual profile by some geometric scaling factor,
$\eta$, explaining why the y-axis of Figure~4 is labeled $\eta f_{0}$.  The
red and green boxes in Figure~4 show the inferred flux overlying Lyman and
Werner band transitions, respectively, which
pump observed fluorescence sequences.
These data points collectively map out a
self-consistent Ly$\alpha$ line profile, albeit with some scatter.
The consistency of the Werner band data points, which are entirely from
the FUSE data, and the Lyman band data points, which are mostly from
the STIS data, provides further evidence that the Ly$\alpha$ profile
is the same at the time of the FUSE and STIS observations.

     Our best estimate for the reconstructed Ly$\alpha$ profile is
indicated by the dashed line in Figure~4.  This is not very different
from the profile derived in Paper~II (also shown in Fig.~4).  The biggest
difference between the two profiles is below 1215.5~\AA, which is
entirely due to the addition of the new 3--1 R(15) Lyman band sequence
fluoresced at 1214.4648~\AA.  This is the only sequence fluoresced at
wavelengths blueward of the stellar rest frame.  Its existence demonstrates
that the H$_{2}$ sees at least some Ly$\alpha$ flux on the blue side
of the line.  As described in Paper~II, the reason the observed and
reconstructed Ly$\alpha$ lines are highly redshifted is because of
absorption from Mira~B's wind on the blue side of the line.  The upper
limits in Figure~4 are from Lyman and Werner band transitions within
Ly$\alpha$ that do not fluoresce enough H$_{2}$ emission to be detectable.
These upper limits are consistent with the Ly$\alpha$ profile derived
from the detected fluorescence sequences.

     Photodissociation rates were overestimated in Paper~II
due to the use of inaccurate dissociation fractions
(see above).  Thus, it is worthwhile to derive better estimates for
H$_{2}$ photodissociation rates using the revised analysis presented here.
We computed these rates for the detected sequences listed in Table~2, using
the $F_{obs}$ and $f_{dis}$ values in that table, and we
found that some of the weaker sequences that are newly detected in our FUSE
data [e.g., 5--0 P(20)] are surprisingly strong contributors
to the total H$_{2}$ dissociation rate.  This suggests that undetected
sequences might collectively contribute substantially to
H$_{2}$ dissociation via Ly$\alpha$ fluorescence.  Therefore, the best
way to accurately estimate the H$_{2}$ photodissociation rate is to
perform a forward modeling exercise computing contributions for {\em all}
H$_{2}$ transitions within Ly$\alpha$, not just the ones that yield
detected H$_{2}$ emission.  The numerous upper limits in Figure~4 provide
some idea for how many H$_{2}$ transitions must be considered.

     We perform this calculation using the Monte Carlo radiative transfer
routine described above and in Paper~II, assuming the Ly$\alpha$ profile
derived in Figure~4, and assuming the $T=3900$~K and $\log N(H_{2})=17.1$
values from the line ratio analysis to compute the H$_{2}$
level populations.  Figure~5 shows a simulated UV H$_{2}$ spectrum
computed from this calculation.  We estimate that the total Ly$\alpha$
flux absorbed and reemitted by H$_{2}$ is $1.06\times 10^{-12}$
ergs~cm$^{-2}$~s$^{-1}$, about 31\% higher than the flux one
would estimate from the detected sequences alone.  About 6.8\% of the
fluorescences lead to H$_{2}$ dissociation rather than the emergence of
an H$_{2}$ line photon, suggesting a dissociation rate of
$8.6\times 10^{39}$ s$^{-1}$ (at Mira's distance of 128~pc),
or $4.5\times 10^{-10}$ M$_{\odot}$~yr$^{-1}$.  Surprisingly, this is
a factor of 2.1 times higher than the value one would derive from the
detected sequences alone.  Although the undetected sequences absorb
less Ly$\alpha$ flux, they tend to have higher dissociation fractions
and therefore contribute significantly to the dissociation rate.
Nevertheless, the largest contributor to H$_{2}$ dissociation is the
detected 4--0 P(19) Lyman band sequence, which accounts for 23.5\% of the
total dissociation rate.

     Although the dissociation fractions assumed in Paper~II were too high,
this is mitigated by the legitimate increase in dissociation
rate provided by consideration of undetected fluorescence sequences.
Thus, the dissociation rate quoted above is only a factor
of 2 lower than reported in Paper~II.  In Paper~II, we proposed that the
H$_{2}$ emission is arising within a photodissociation front within
Mira~A's wind as it approaches Mira~B, and we constructed a model of
this front consistent with the observations.  We experiment with
new photodissociation front models assuming our revised estimates for
photodissociation rates, and we find that the factor of 2 change is not
enough to significantly alter the model.  Thus, the photodissociation
front model remains a very viable interpretation for the H$_{2}$ emission.
As mentioned in \S 1, the implied H$_{2}$ dissociation rate is comparable
to Mira~B's total accretion rate, consistent with the idea that the UV
H$_{2}$ emission is showing us H$_{2}$ in Mira~A's wind being fluoresced,
heated, and dissociated by Ly$\alpha$ as it approaches and is ultimately
accreted by Mira~B.

     We now discuss some of the limitations of our H$_{2}$
modeling efforts, because our estimation of $T$ and $N(H_{2})$ and our
reconstruction of the Ly$\alpha$ profile in Figure~4 rely on several
assumptions.  For example, our plane-parallel modeling essentially
assumes that Mira~B is surrounded by an isothermal H$_{2}$ layer of
uniform $N(H_{2})$ and that H$_{2}$ level populations are precisely
thermal.  In reality, the H$_{2}$ may have a distribution of temperatures
(as the photodissociation front models actually predict), it may be
distributed inhomogenously around the star, and the fluorescence process
itself could lead to some level of nonthermality in the level populations.
The limitations introduced by our simplifying assumptions are presumably
responsible for the significant scatter of Ly$\alpha$ flux data points in
Figure~4 about the best-fit dashed line Ly$\alpha$ profile, beyond the
measurement uncertainties.  Since the reconstructed profile imprecisely
reproduces the data points in Figure~4, the H$_{2}$ fluxes of the synthetic
spectrum described above are naturally imprecise as well.  For example,
the synthetic spectrum predicts lines at 1130~\AA\ and 1135~\AA\ that
are significantly stronger than observed and a line at 1162~\AA\
significantly weaker than observed (see Fig.~5).  Another source of
uncertainty is line blending, which is not considered in our
radiative transfer calculations.  For example, there are 2 Lyman band data
points at 1219~\AA\ in Figure~4 that are too low, probably because these
H$_{2}$ pumping transitions are blended enough that they are partially
shielding each other from the full Ly$\alpha$ flux present at that
wavelength.  Emission line blends could also lead to H$_{2}$ photons
shifting from one fluorescence sequence to another.  All the effects
and assumptions described above could lead to inaccuracies in our derived
H$_{2}$ properties.

\section{THE C III LINES AND MIRA B'S VARIABILITY}

     One goal of our FUSE observations was to try to detect broad, high
temperature emission lines formed in the accretion disk of Mira~B.
We see no evidence of the O~VI $\lambda\lambda$1032,1038 lines, which are
typically very strong lines in coronal spectra.  We estimate 
a 3$\sigma$ upper limit of $9\times 10^{-16}$ ergs~cm$^{-2}$~s$^{-1}$ for
these lines.  The only possible
accretion lines are the two C~III lines shown in Figure~1.  The widths
of these two lines are roughly consistent with observations of high
temperature lines from IUE \citep{dr85}, but the
identification with C~III is still somewhat tentative.  We
identify the broad bump at 976.447~\AA\ as C~III $\lambda$977, but
if it is C~III it has a very large blueshift of $-232$ km~s$^{-1}$,
assuming a stellar velocity of $+56$ km~s$^{-1}$
\citep{pfb88,pp90,ej00}.
The scattered solar C~III $\lambda$977 line is sometimes seen in FUSE
spectra, but the observed feature is far too broad for that to be
responsible.  The spike seen on the red edge of the line could perhaps
be the solar emission.

     One possible explanation for the blueshift of the C~III line is
that we are seeing emission predominantly from the side of the accretion
disk that is rotating towards us.  There is no corresponding
blueshift of the C~III $\lambda$1175 line (see below), but the $\lambda$977
line is a high opacity resonance line, meaning intervening C~III material
could in principle scatter background $\lambda$977 emission, while the
$\lambda$1175 lines are lower opacity intersystem lines, so the same process
may not work for $\lambda$1175.  Another possible explanation for the C~III
blueshift is absorption from overlying H$_{2}$.  The density of H$_{2}$
transitions increases towards the shorter wavelengths accessible to FUSE.
These transitions will have higher opacities than any of the observed
H$_{2}$ lines, since they will be originating from lower, more populated
levels of the ground electronic state.  Below about 1110~\AA\ there are
Lyman and Werner band transitions from the lowest levels that will be
populated even at very cool temperatures.  The column densities of H$_{2}$
in these lowest levels could be much higher than for the $T=3900$~K
population that we have detected via Ly$\alpha$ fluorescence.  This is
because most of the H$_{2}$ within Mira~A's wind, in which Mira~B is
embedded, will be at low $T\sim 100$~K temperatures, and only the lowest
energy levels will be populated in this cold H$_{2}$.  Because of the
significant H$_{2}$ opacity in the neighborhood of the C~III $\lambda$977
line, the red side of the line might be absorbed, making the resulting
C~III profile appear as blueshifted as seen in our FUSE data.  Despite
these plausible explanations for the blueshift, the detection of the
$\lambda$977 line will remain tentative until a more definitive
explanation of the blueshift is available.

     The C~III $\lambda$1175 multiplet is blended with many H$_{2}$
emission lines and it is therefore unclear that C~III is actually
contributing to the observed emission.
At the end of \S4 we described a forward modeling calculation designed
to compute the total H$_{2}$ photodissociation rate from Ly$\alpha$
fluorescence, including contributions from undetected fluorescence
sequences.  This calculation also provides us with the synthetic H$_{2}$
emission spectrum shown in Figure~5.  This figure compares the
synthetic spectrum with the FUSE data, demonstrating a reasonably good fit
to the observed H$_{2}$ lines.  The bottom panel focuses on the
1176~\AA\ region.  We subtract the H$_{2}$ spectrum from the data to see
if the H$_{2}$ lines can account for all of the line flux.  The residual
flux shown below the lowest panel of Figure~5 suggests that there is
flux remaining after this subtraction that we can associate with
the C~III $\lambda$1175 multiplet.  The integrated flux remaining is
$1.4\times 10^{-14}$ ergs~cm$^{-2}$~s$^{-1}$, in good agreement with
the C~III flux suggested by the two-Gaussian representation in
Figure~2, providing support for this being a reasonably
accurate C~III flux measurement.  The flux ratio of C~III $\lambda$977 and
C~III $\lambda$1175 is often used as a density diagnostic, but in Mira~B's
case the significant uncertainties involved in measuring both lines are
too large to obtain a believable density measurement.

     The 1176~\AA\ emission feature is the only feature in our FUSE
spectrum that is also detectable in the previous HST/STIS spectrum
(see Fig.~2 in Paper I).  The sensitivity of STIS drops dramatically
below 1200~\AA, so the S/N of the STIS spectrum at 1176~\AA\ is very
low.  The data quality does not allow us to separate the H$_{2}$ emission
from the C~III emission, so we can only report a total STIS flux of
$(2.7\pm 0.8)\times 10^{-14}$ ergs~cm$^{-2}$~s$^{-1}$ for the
1176~\AA\ line.  This agrees well with the integrated FUSE flux
of $(3.00\pm 0.13)\times 10^{-14}$ ergs~cm$^{-2}$~s$^{-1}$.
In \S3 and \S4, we demonstrated that the H~I Ly$\alpha$ line had not
changed between the times of the STIS and FUSE observations.  The
excellent flux agreement for the 1176~\AA\ feature provides further
evidence that the UV spectrum of Mira~B is unchanged.

     The C~III $\lambda$1175 line has also been observed numerous times
by IUE.  We found 32 short-wavelength, low-resolution (SW-LO) observations
of Mira~B in the IUE archive.  Only about half of them are long enough
exposures with sufficient S/N to measure a C~III flux.  Note that while
H$_{2}$ may be contaminating the C~III line in the HST/STIS and FUSE
data, we do not believe that this is the case for the IUE data.  In Paper~I
it was argued that unblended H$_{2}$ features should have been
seen in the IUE data if H$_{2}$ were contributing flux to other emission
features such as C~III $\lambda$1175.

     In Figure~6, we plot the C~III fluxes as a function of time, where
for purposes of this figure we have made no attempt to correct the STIS
and FUSE fluxes for H$_{2}$ contamination.  The STIS and FUSE fluxes are
about an order of magnitude lower than the IUE fluxes, consistent with
the drop in flux seen in the continuum and many other emission lines (see
Paper~I).  The drop in flux is presumably due to a substantial decrease in
accretion rate, which also led to a similar decrease in mass loss rate for
Mira~B (see Paper~II).  The consistency of the 1999 STIS and
2001 FUSE data demonstrates that this decrease is an enduring phenomenon.

     It is possible that similar changes in accretion rate have been
detected from optical data.  Mira~B is difficult to observe from the
ground due to its close proximity with the generally brighter Mira~A, but
during Mira~A minimum it is possible to detect the presence of Mira~B.
\citet{ahj54} and \citet{yy77} report a possible periodic
variation in Mira~B's optical light curve with a period of 14 years and
with an intensity range of about a factor of 4.  Considering that
these optical intensities will be contaminated with Mira~A emission even
at Mira~A minimum, these variations might be consistent with the factor
of 20 variations seen in the UV.

     However, if the period of these variations is 14 years, why did the
long-lived IUE not see them?  \citet{yy77} report a
Mira~B optical minimum in 1971, so in Figure~6 we display a 14-year
period sine curve consistent with this phasing.  All the IUE data points
occur near predicted maxima, while the STIS and FUSE data points are
near a predicted minimum.  There is an unfortunate time gap in the IUE
data from 1981--1990 that means the predicted Mira~B minimum period was
not covered by the IUE SW-LO data set.  Thus, it is possible that the
substantial drop in flux seen by STIS and FUSE is simply the UV
manifestation of the 14-year optical cycle detected by \citet{ahj54} and
\citet{yy77}.  One argument against this is that while there
are no SW-LO IUE observations from 1981--1990, there are long-wavelength,
low-resolution (LW-LO) spectra from 1983 July 9 and 1988 January 8.
Although Mg~II $\lambda$2800 and continuum fluxes are somewhat lower than
average in these spectra, neither spectrum shows flux levels dramatically
below those of other LW-LO observations in the IUE data set.  This leaves
only about a 5-year time gap in the IUE coverage during which Mira~B fluxes
could have dropped tremendously like they did in 1999--2001.  Nevertheless,
it would be worthwhile to observe Mira~B again with FUSE and/or HST in
2004--2007 to see if high UV fluxes return as Figure~6 predicts.

\section{SUMMARY}

     We have analyzed UV observations of the wind-accreting star Mira~B,
using new FUSE spectra combined with previous HST/STIS data.  Our results
are summarized as follows:
\begin{description}
\item[1.] In the new FUSE data, we detect Ly$\alpha$-fluoresced H$_{2}$
  lines that are mostly Werner band lines rather than the Lyman band lines
  previously detected by HST.  Most of the FUSE H$_{2}$ lines have never
  before been identified in an astrophysical spectrum.
\item[2.] Using previously developed techniques, we analyze the Mira~B
  H$_{2}$ emission, combining the old HST/STIS and the new FUSE H$_{2}$ data.
  We estimate a temperature and column density for the H$_{2}$ layer
  responsible for the emission of $T=3900$~K and $\log N(H_{2})=17.1$,
  respectively.
\item[3.] Our modeling efforts demonstrate that undetected H$_{2}$
  fluorescence sequences actually
  produce more H$_{2}$ photodissociation from Ly$\alpha$ fluorescence than
  do the detected sequences.  Considering both, we estimate a total
  photodissociation rate of $4.5\times 10^{-10}$ M$_{\odot}$~yr$^{-1}$,
  comparable to the $(8-30)\times 10^{-10}$ M$_{\odot}$ yr$^{-1}$ total
  accretion rate of Mira~B.  The FUSE and HST H$_{2}$ emission may be
  coming from a photodissociation front, as first proposed in the analysis
  of the HST data (see Paper~II).
\item[4.] The only stellar lines detected in our FUSE spectrum other
  than H$_{2}$ are the broad C~III $\lambda$977 and C~III $\lambda$1175
  lines, which originate from within Mira~B's accretion disk.  The C~III
  $\lambda$977 line is highly blueshifted from its expected location,
  which leads us to consider its identification as tentative.  Overlying
  H$_{2}$ absorption or particularly bright emission from one side of the
  accretion disk could in principle be responsible for the blueshift.
  The C~III $\lambda$1175 line is heavily blended with numerous H$_{2}$
  emission lines, making an accurate flux measurement difficult.
\item[5.] Analysis of the H$_{2}$ and C~III $\lambda$1175 lines
  in the 1999 HST/STIS and 2001 FUSE data demonstrates that the UV spectrum
  of Mira~B is roughly the same at these two times.  The UV fluxes of
  both data sets are dramatically lower than ever observed by IUE in
  1979--1995.  The presence of the lower fluxes over at least two years
  (1999--2001) demonstrates that they are a persistent phenomenon.
\item[6.] We hypothesize that the drop in UV flux in 1999--2001 is
  associated with a previously identified 14-year periodic variation in
  optical emission from Mira~B, and that IUE missed the variation due
  to most IUE observations falling near the two maxima of the
  cycle during the IUE era.
\end{description}

\acknowledgments

Support for this work was provided by NASA through grant NAG5-11950
to the University of Colorado.  M.\ K.\ is a member of the
Chandra Science Center, which is operated under contract NAS8-39073,
and is partially supported by NASA.

\clearpage

\begin{deluxetable}{ccclccl}
\tabletypesize{\scriptsize}
\tablecaption{Line List}
\tablecolumns{7}
\tablewidth{0pt}
\tablehead{
  \colhead{$\lambda_{meas}$} & \colhead{$f_{obs}$} & \colhead{FWHM} &
    \colhead{ID} & \colhead{$\lambda_{rest}$} &
    \colhead{$f_{mod}$\tablenotemark{a}} &
    \colhead{Fluoresced by} \\
  \colhead{(\AA)}            &\colhead{($10^{-15}$)}&\colhead{(km s$^{-1}$)}&
    \colhead{}   & \colhead{(\AA)}           &\colhead{($10^{-15}$)}&
    \colhead{}}
\startdata
\multicolumn{7}{l}{\underline{FUSE H$_{2}$ Lines}} \\
$1127.433\pm 0.008$ & $2.87\pm 0.23$ &  $50\pm 5$  & 0-2 Q(10) &
  1127.160 & 1.60 & 0-4 Q(10) $\lambda$1217.263 \\
                    &                &             & 1-3 P(5)  &
  1127.245 & 0.78 & 1-5 P(5) $\lambda$1216.988 \\
$1130.571\pm 0.015$ & $0.66\pm 0.14$ &  $31\pm 7$  & 1-3 Q(7)  &
  1130.365 & 1.65 & 1-5 Q(7) $\lambda$1218.507 \\
$1131.608\pm 0.015$ & $0.98\pm 0.18$ &  $37\pm 8$  & 2-4 R(1)  &
  1131.309 & 0.22 & 2-6 R(1) $\lambda$1217.298 \\
                    &                &             & 1-3 R(9)  &
  1131.390 & 1.03 & 1-5 R(9) $\lambda$1216.997 \\
$1143.257\pm 0.021$ & $0.61\pm 0.16$ &  $39\pm 9$  & 2-4 P(5)  &
  1142.956 & 0.38 & 2-6 R(3) $\lambda$1217.488 \\
                    &                &           &2-0 P(11)\tablenotemark{b}&
  1143.112 & 0.07 & 2-2 R(9) $\lambda$1219.101 \\
$1155.154\pm 0.008$ & $1.53\pm 0.23$ &  $24\pm 4$  & 1-3 P(11) &
  1154.910 & 1.84 & 1-5 R(9) $\lambda$1216.997 \\
$1160.912\pm 0.010$ & $1.01\pm 0.20$ &  $24\pm 5$  & 1-4 R(3)  &
  1160.647 & 1.38 & 1-5 P(5) $\lambda$1216.988 \\
$1162.051\pm 0.019$ & $2.18\pm 0.32$ &  $61\pm 10$&0-1 R(0)\tablenotemark{b}&
  1161.693 & 0.72 & 0-2 R(0) $\lambda$1217.205 \\
                    &                &            &1-1 P(5)\tablenotemark{b}&
  1161.814 & 0.16 & 1-2 P(5) $\lambda$1216.070 \\
                    &                &            &1-1 R(6)\tablenotemark{b}&
  1161.949 & 0.21 & 1-2 R(6) $\lambda$1215.726 \\
$1172.242\pm 0.006$ & $4.60\pm 0.40$ &  $41\pm 4$  & 1-4 P(5)  &
  1171.953 & 1.62 & 1-5 P(5) $\lambda$1216.988 \\
                    &                &             & 0-3 Q(10) &
  1172.031 & 2.67 & 0-4 Q(10) $\lambda$1217.263 \\
$1174.571\pm 0.007$ & $4.29\pm 0.50$ &  $37\pm 5$  & 2-5 R(1)  &
  1174.298 & 1.41 & 2-6 R(1) $\lambda$1217.298 \\
                    &                &             & 1-4 Q(7)  &
  1174.347 & 4.04 & 1-5 Q(7) $\lambda$1218.507 \\
$1174.846\pm 0.013$ & $1.25\pm 0.35$ &  $28\pm 9$  & 2-5 R(0)  &
  1174.574 & 0.29 & 2-6 R(0) $\lambda$1217.680 \\
                    &                &             & 2-5 R(2)  &
  1174.588 & 0.30 & 2-6 R(2) $\lambda$1217.440 \\
$1175.110\pm 0.007$ & $2.07\pm 0.35$ &  $25\pm 4$  & 2-5 R(3)  &
  1174.869 & 1.64 & 2-6 R(3) $\lambda$1217.488 \\
$1176.060\pm 0.013$ & $0.89\pm 0.28$ &  $25\pm 8$  & 2-5 Q(1)  &
  1175.826 & 0.78 & 2-6 Q(1) $\lambda$1218.940 \\
$1176.319\pm 0.010$ & $2.22\pm 0.36$ &  $34\pm 6$&5-0 R(18)\tablenotemark{b}&
  1176.066 & 0.57 & 5-0 P(20) $\lambda$1217.717 \\
                    &                &             & 0-2 Q(18) &
  1176.082 & 1.58 & 0-3 Q(18) $\lambda$1216.692 \\
                    &                &           & 2-5 R(4)  &
  1176.085 & 0.33 & 2-6 R(4) $\lambda$1218.457 \\
$1176.978\pm 0.016$ & $0.60\pm 0.25$ & $20\pm 10$&6-0 R(19)\tablenotemark{b}&
  1176.706 & 0.53 & 6-0 P(21) $\lambda$1218.841 \\
                    &                &           & 2-5 Q(2)  &
  1176.788 & 0.19 & 2-6 Q(2) $\lambda$1219.804 \\
$1177.518\pm 0.026$ & $0.95\pm 0.36$ &  $39\pm 17$ & 1-3 Q(16) &
  1177.269 & 1.75 & 1-4 Q(16) $\lambda$1216.930 \\
$1180.710\pm 0.013$ & $0.64\pm 0.18$ &  $16\pm 7$  & 2-5 P(3)  &
  1180.457 & 1.38 & 2-6 R(1) $\lambda$1217.298 \\
\multicolumn{7}{l}{\underline{Other FUSE Lines}} \\
 $976.447\pm 0.121$ & $6.99\pm 1.60$ & $338\pm 84$ & C III?    &
   977.020 &   &  \\
$1174.616\pm 0.083$ & $7.03\pm 1.82$ & $287\pm 47$ & C III?    &
  1175mult &   &  \\
$1177.206\pm 0.115$ & $8.57\pm 1.32$ & $405\pm 68$ & C III?    &
  1175mult &   &  \\
\multicolumn{7}{l}{\underline{New STIS H$_{2}$ Lines}} \\
$1254.369\pm 0.013$ & $1.01\pm 0.31$ &  $16\pm 4$&3-1 P(17)\tablenotemark{b}&
  1254.125 & 0.26 & 3-1 R(15) $\lambda$1214.465 \\
$1265.421\pm 0.009$ & $0.99\pm 0.27$ &  $17\pm 5$&3-2 R(15)\tablenotemark{b}&
  1265.180 & 0.31 & 3-1 R(15) $\lambda$1214.465 \\
$1593.575\pm 0.016$ & $3.77\pm 0.83$ &  $23\pm 6$&3-9 R(15)\tablenotemark{b}&
  1593.258 & 0.74 & 3-1 R(15) $\lambda$1214.465 \\
$1621.385\pm 0.021$ & $3.54\pm 0.89$ & $26\pm 8$&3-10 R(15)\tablenotemark{b}&
  1621.119 & 0.39 & 3-1 R(15) $\lambda$1214.465 \\
$1622.434\pm 0.011$ & $5.19\pm 1.01$ &  $26\pm 6$&3-9 P(17)\tablenotemark{b}&
  1622.133 & 0.87 & 3-1 R(15) $\lambda$1214.465 \\
\enddata
\tablenotetext{a}{Fluxes predicted using the Ly$\alpha$ profile and
  H$_{2}$ properties from Paper II.}
\tablenotetext{b}{A Lyman band line.  All other H$_{2}$ lines are Werner
  band.}
\end{deluxetable}

\clearpage

\begin{deluxetable}{cccccccc}
\tabletypesize{\small}
\tablecaption{Fluorescence Sequences}
\tablecolumns{8}
\tablewidth{0pt}
\tablehead{
  \colhead{Fluorescing} & \colhead{$\lambda_{rest}$} & \colhead{$F_{obs}$} &
    \colhead{Absorption} & \colhead{$E_{low}$} & \colhead{$g_{low}$} &
    \colhead{$f_{dis}^{\prime}$} & \colhead{$f_{dis}$} \\
  \colhead{Transition} & \colhead{(\AA)} & \colhead{($10^{-13}$)} & 
    \colhead{Strength (f)} & \colhead{(cm$^{-1}$)} & \colhead{} & \colhead{}}
\startdata
\multicolumn{8}{l}{\underline{Lyman Band Sequences from Paper II}} \\
1-2 R(6)  & 1215.7263 & 0.502 & 0.0349 & 10261.20 &  13 & 0.0 & 0.0 \\
1-2 P(5)  & 1216.0696 & 0.914 & 0.0289 &  9654.15 &  33 & 0.0 & 0.0 \\
3-3 P(1)  & 1217.0377 & 0.115 & 0.0013 & 11883.51 &   9 & 0.0 & 0.0 \\
0-2 R(0)  & 1217.2045 & 0.858 & 0.0441 &  8086.93 &   1 & 0.0 & 0.0 \\
4-0 P(19) & 1217.4100 & 0.190 & 0.0093 & 17750.25 & 117 & 0.417 & 0.469 \\
0-2 R(1)  & 1217.6426 & 1.355 & 0.0289 &  8193.81 &   9 & 0.0 & 0.0 \\
2-1 P(13) & 1217.9041 & 1.141 & 0.0192 & 13191.06 &  81 & 0.002 & 0.002 \\
2-1 R(14) & 1218.5205 & 0.324 & 0.0181 & 14399.08 &  29 & 0.006 & 0.006 \\
0-2 R(2)  & 1219.0887 & 0.165 & 0.0256 &  8406.29 &   5 & 0.0 & 0.0 \\
2-2 R(9)  & 1219.1005 & 0.320 & 0.0318 & 12584.80 &  57 & 0.0 & 0.0 \\
2-2 P(8)  & 1219.1543 & 0.340 & 0.0214 & 11732.12 &  17 & 0.0 & 0.0 \\
0-2 P(1)  & 1219.3676 & 0.318 & 0.0149 &  8193.81 &   9 & 0.0 & 0.0 \\
0-1 R(11) & 1219.7454 & 0.166 & 0.0037 & 10927.12 &  69 & 0.0 & 0.0 \\
\multicolumn{8}{l}{\underline{New Lyman Band Sequences}} \\
3-1 R(15) & 1214.4648 & 0.284 & 0.0236 & 15649.58 &  93 & 0.031 & 0.037 \\
5-0 P(20) & 1217.7165 & 0.069 & 0.0115 & 19213.16 &  41 & 0.478 & 0.499 \\
6-0 P(21) & 1218.8403 & 0.051 & 0.0130 & 20688.04 & 129 & 0.520 & 0.555 \\
\multicolumn{8}{l}{\underline{Werner Band Sequences}} \\
0-3 Q(18) & 1216.6926 & 0.036 & 0.0396 & 25499.74 &  37 & 0.0 & 0.0 \\
1-4 Q(16) & 1216.9302 & 0.034 & 0.0712 & 25929.42 &  33 & 0.0 & 0.0 \\
1-5 P(5)  & 1216.9878 & 0.072 & 0.0071 & 19807.03 &  33 & 0.016 & 0.031 \\
1-5 R(9)  & 1216.9969 & 0.099 & 0.0197 & 22251.21 &  57 & 0.017 & 0.031 \\
0-4 Q(10) & 1217.2628 & 0.065 & 0.0100 & 20074.45 &  21 & 0.0 & 0.0 \\
2-6 R(1)  & 1217.2978 & 0.052 & 0.0558 & 21589.82 &   9 & 0.008 & 0.012 \\
2-6 R(2)  & 1217.4400 & 0.047 & 0.0399 & 21756.98 &   5 & 0.066 & 0.089 \\
2-6 R(3)  & 1217.4883 & 0.090 & 0.0364 & 22005.58 &  21 & 0.073 & 0.113 \\
2-6 R(0)  & 1217.6800 & 0.027 & 0.1098 & 21505.78 &   1 & 0.002 & 0.002 \\
2-6 R(4)  & 1218.4566 & 0.015 & 0.0399 & 22332.85 &   9 & 0.017 & 0.025 \\
1-5 Q(7)  & 1218.5084 & 0.055 & 0.0303 & 20894.94 &  45 & 0.0 & 0.0 \\
2-6 Q(1)  & 1218.9402 & 0.023 & 0.0532 & 21589.82 &   9 & 0.0 & 0.0 \\
2-6 Q(2)  & 1219.8038 & 0.004 & 0.0532 & 21756.98 &   5 & 0.0 & 0.0 \\
\enddata
\end{deluxetable}

\clearpage

\begin{figure}
\plotfiddle{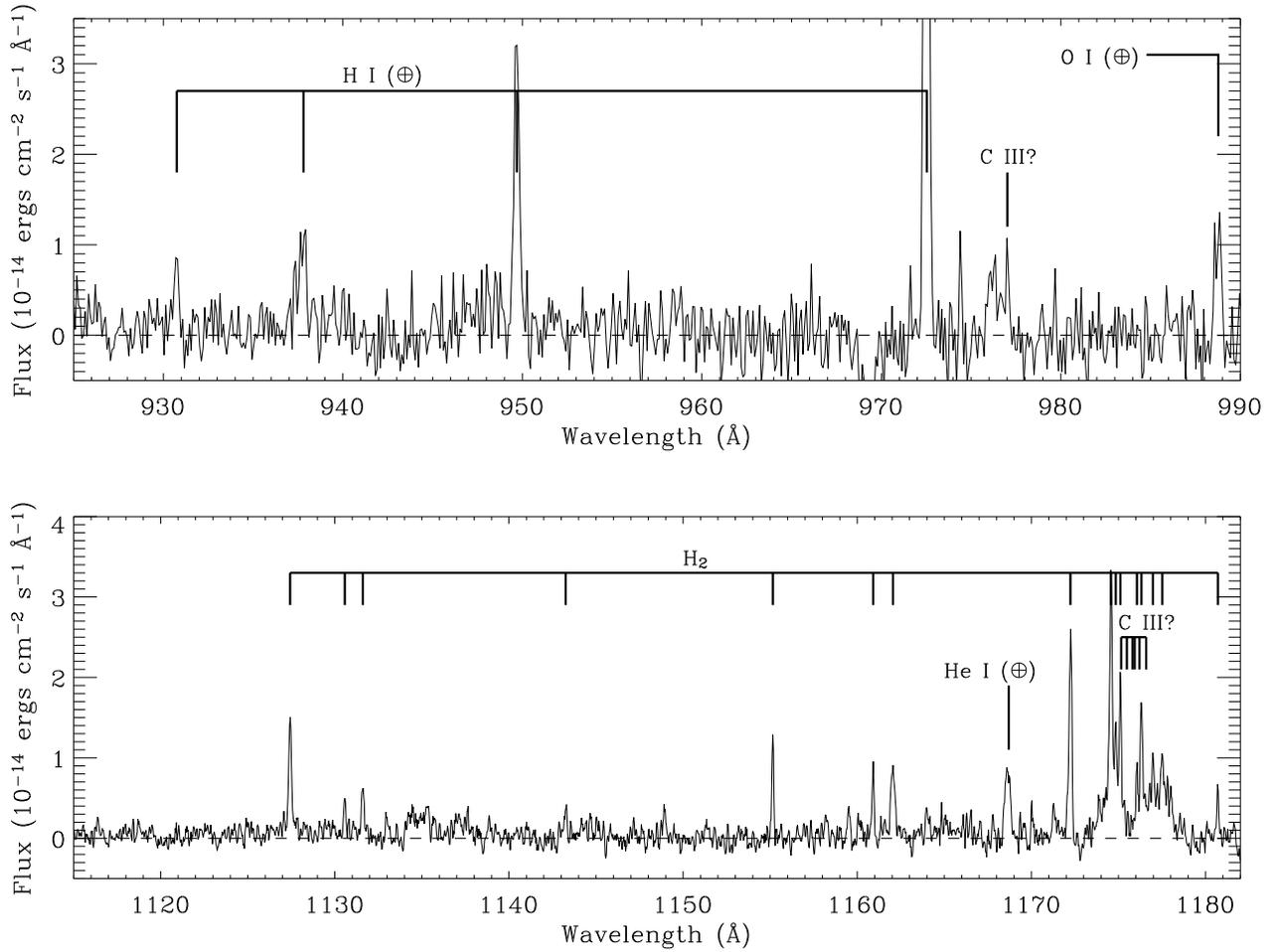}{3.5in}{90}{75}{75}{295}{0}
\caption{Two regions of the FUSE spectrum of Mira~B.  Line identifications
  are displayed, where the $\oplus$ symbol indicates a line is an airglow
  line rather than a stellar feature.}
\end{figure}

\clearpage

\begin{figure}
\plotfiddle{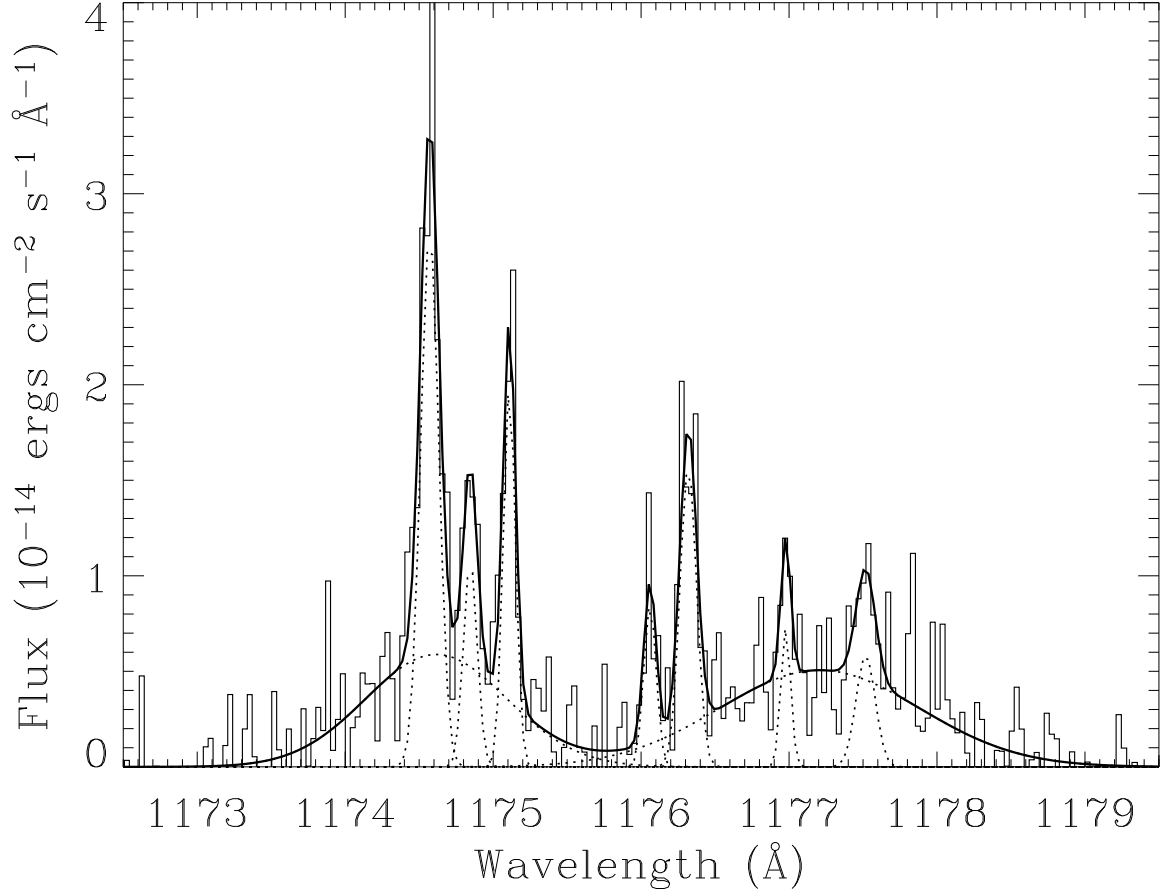}{3.5in}{90}{75}{75}{290}{0}
\caption{Multi-Gaussian fit to the 1176~\AA\ emission feature.  The
  7 narrow components are H$_{2}$ lines, and we assume the sum of the
  2 broad components is representative of the C~III $\lambda$1175
  line profile.}
\end{figure}

\clearpage

\begin{figure}
\plotfiddle{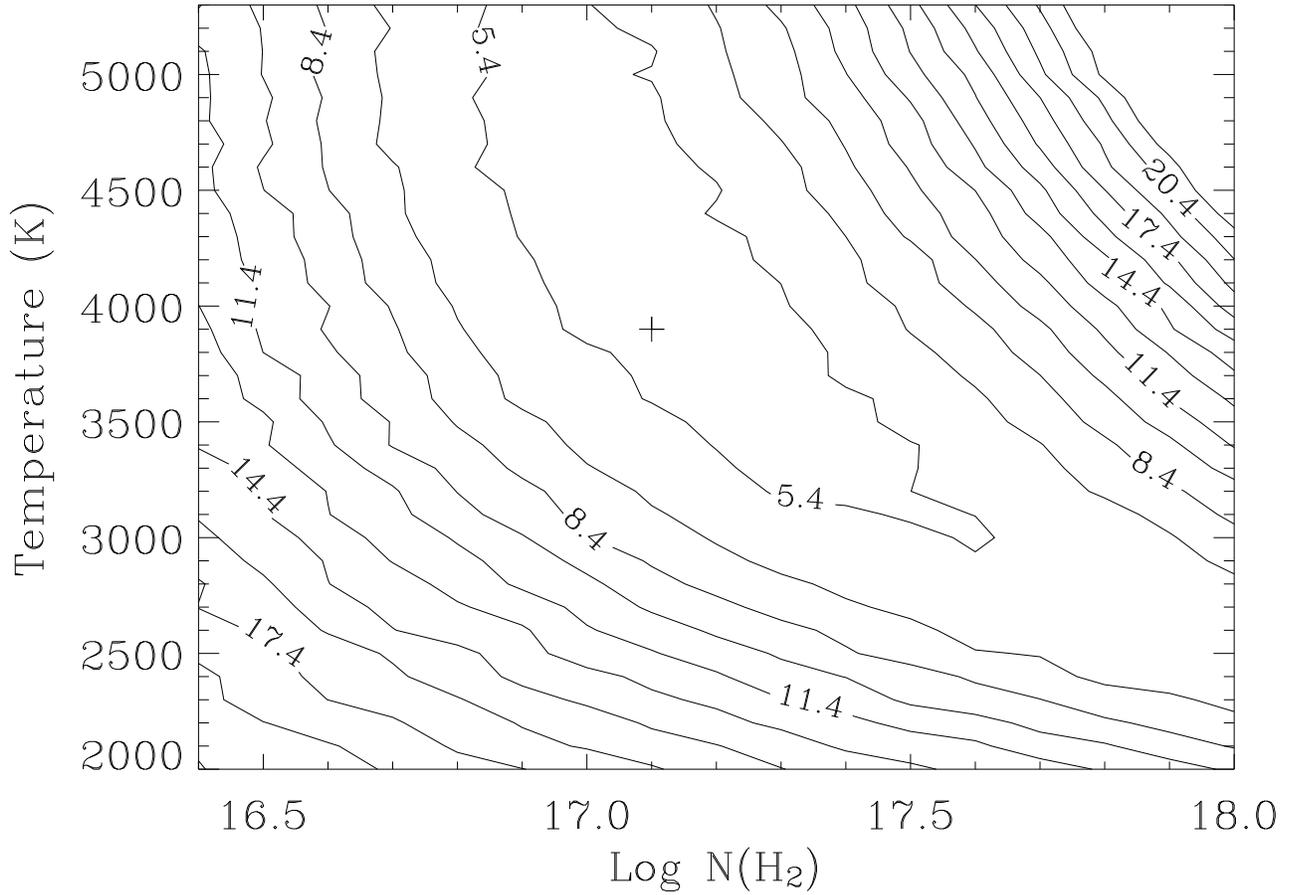}{3.5in}{90}{75}{75}{295}{0}
\caption{Reduced $\chi^{2}$ contours measuring how well the H$_{2}$ lines
  ratios can be reproduced as a function of assumed temperature and column
  density.  Both FUSE and HST/STIS H$_{2}$ lines are considered in the
  analysis.  The best fit is at $T=3900$~K and $\log N(H_{2})=17.1$ (plus
  sign).}
\end{figure}

\clearpage

\begin{figure}
\plotfiddle{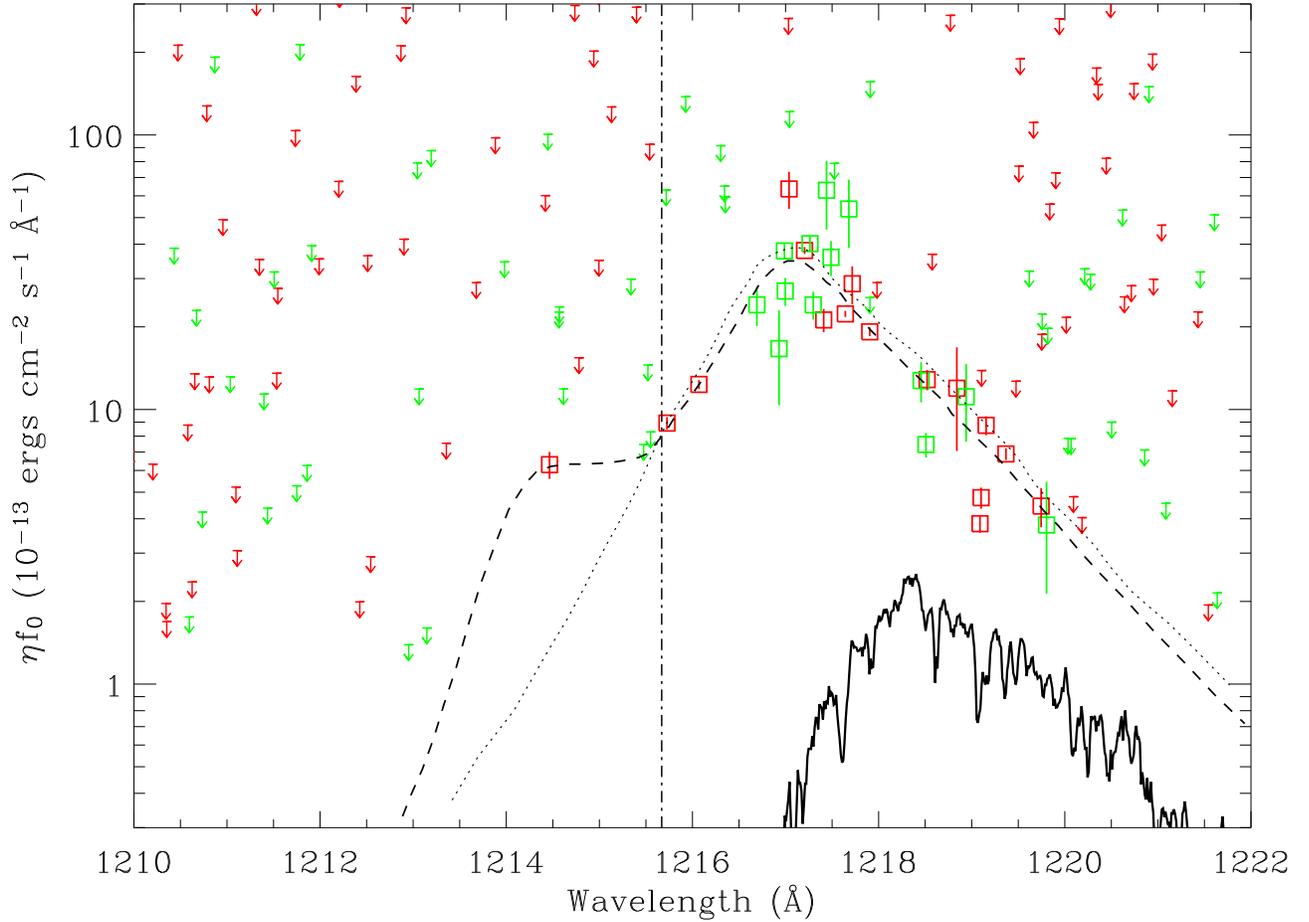}{3.5in}{90}{75}{75}{295}{0}
\caption{Reconstruction of the H~I Ly$\alpha$ profile seen by the
  fluoresced H$_{2}$ surrounding Mira~B.  The boxes are the inferred
  Ly$\alpha$ fluxes overlying Lyman band (red) and Werner band (green)
  H$_{2}$ pumping transitions, which fluoresce observed sequences of H$_{2}$
  lines.  Error bars indicate the 1$\sigma$ uncertainty in the fluxes due
  to uncertainties in the measurements of the H$_{2}$ emission lines.
  Upper limits are shown for undetected sequences.  The dashed line
  is our best estimate for the Ly$\alpha$ profile, which is compared with
  the profile derived from Paper~II (dotted line).  The solid line is the
  actual Ly$\alpha$ line observed by HST/STIS, and the dot-dashed line
  indicates the rest frame of the star.}
\end{figure}

\clearpage

\begin{figure}
\plotfiddle{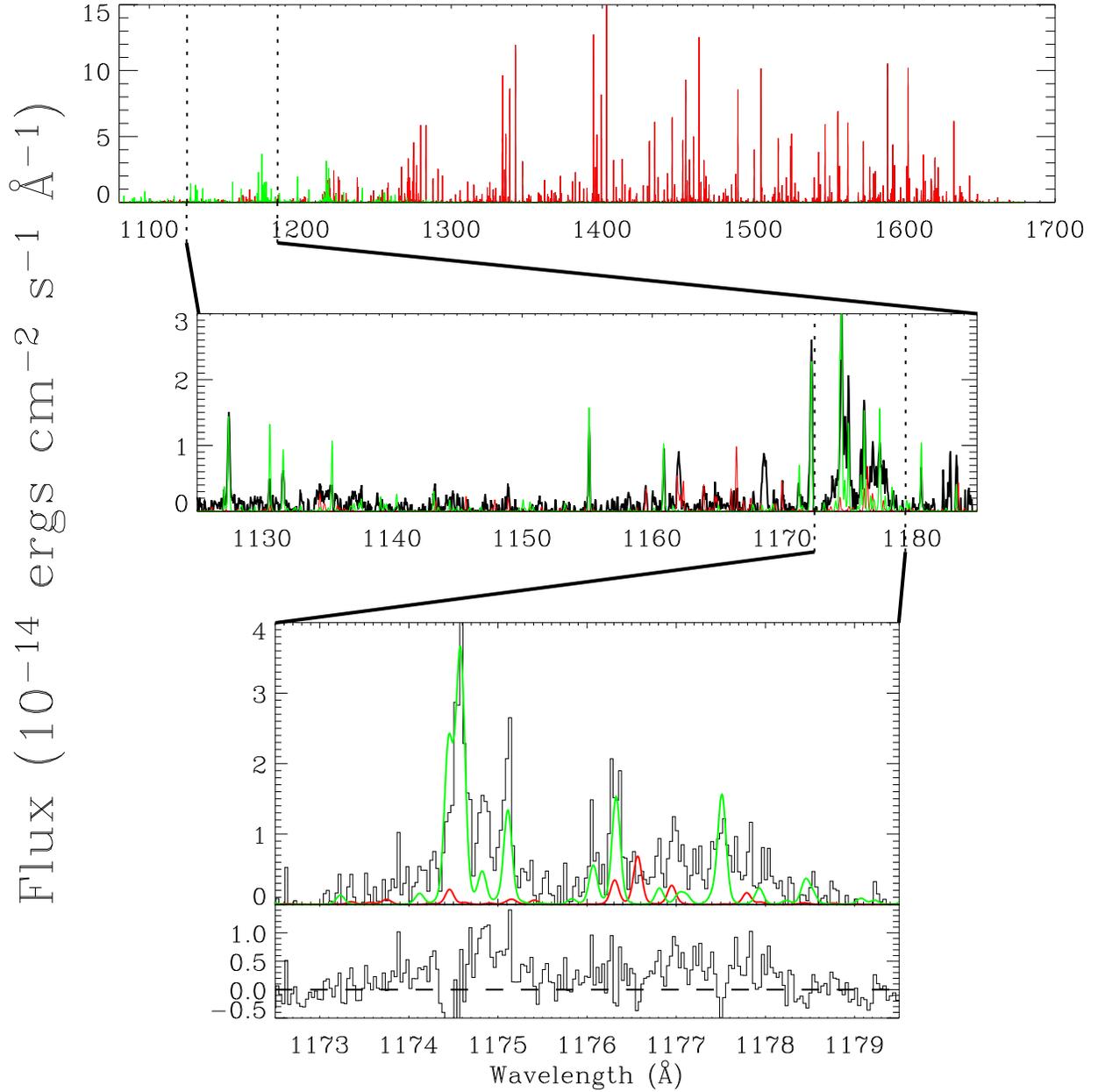}{6.5in}{0}{95}{95}{-315}{10}
\caption{A simulated H$_{2}$ spectrum fluoresced by the Ly$\alpha$ line
  of Mira~B (top panel), where the red and green lines are for Lyman
  and Werner band lines, respectively.  In the lower two panels, this
  simulated spectrum is compared with the FUSE spectrum of Mira~B.
  The bottom panel zooms in on the 1176~\AA\ emission feature, and the
  residuals of the data after subtraction of the model
  spectrum are shown.  The generally positive flux remaining after
  subtraction suggests that H$_{2}$ is not responsible for all the flux
  of the 1176~\AA\ feature, and that broad C~III $\lambda$1175 emission
  is contributing to the line.}
\end{figure}

\clearpage

\begin{figure}
\plotfiddle{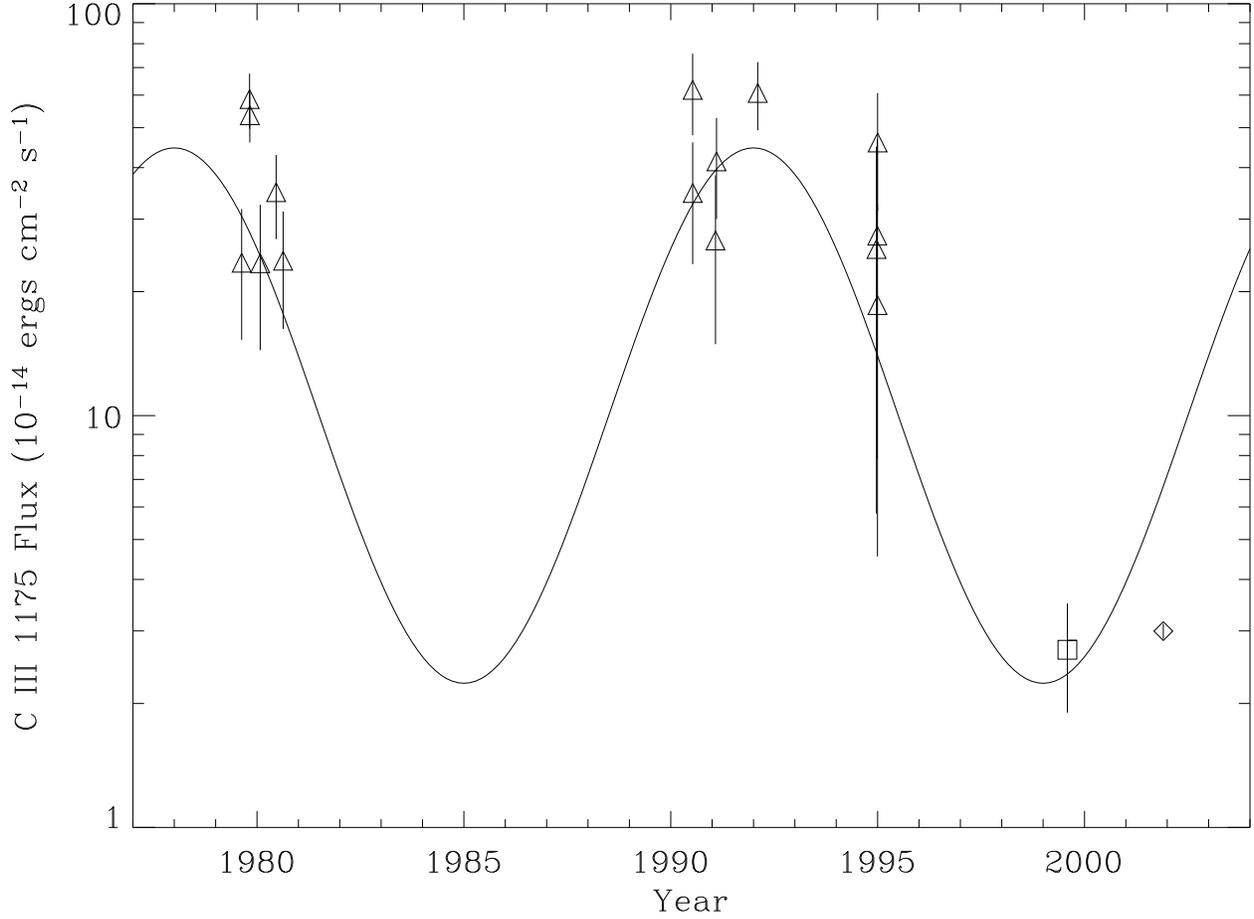}{3.5in}{90}{75}{75}{295}{0}
\caption{C~III $\lambda$1175 flux plotted as a function of time, where
  triangles are IUE measurements, the box is from the HST/STIS spectrum,
  and the diamond is from the FUSE data.  Note that for the HST and FUSE
  data, the C~III feature is heavily contaminated with H$_{2}$ lines, while
  this is probably not the case for the IUE data.  The sine curve is a
  schematic representation of the 14 year periodicity for Mira~B suggested
  by optical observations (Joy 1954; Yamashita \& Maehara 1977).}
\end{figure}

\end{document}